\documentstyle[epsf]{mn}

\onecolumn 

\title{The lower boundary of the accretion column in magnetic cataclysmic 
  variables}

\author[Wu \& Cropper]{ 
Kinwah Wu$^{1,2}$ and Mark Cropper$^2$\\ 
$^1$ Research Centre for Theoretical Astrophysics, School of Physics A28, 
     University of Sydney, NSW 2006, Australia \\
$^2$ Mullard Space Science Laboratory, University College London, 
     Holmbury St Mary, Dorking, Surrey RH5 6NT  }

\date{Received: }

\begin{document}

\def\Mdot{\hbox{$\dot M$}}
\def\Msun{\hbox{M$_\odot$}}
\def\Rwd{\hbox{$R_{_{\rm WD}}$\,}}
\def\GS{\lower.5ex\hbox{$\buildrel>\over\sim$}}
\def\LS{\lower.5ex\hbox{$\buildrel<\over\sim$}}
\def\kms{km ${\rm s}^{-1}$}    

\maketitle

\begin{abstract}
Using a parameterised function for the mass loss 
  at the base of the post-shock region, 
  we have constructed a formulation for magnetically confined accretion flows 
  which avoids singularities, such as the infinity in density, 
  at the base associated with all previous formulations. 
With the further inclusion of a term allowing 
  for the heat input into the base from the accreting white dwarf 
  we are able also to obtain the hydrodynamic variables 
  to match the conditions in the stellar atmosphere. 
(We do not, however, carry out a mutually consistent analysis for the match). 
  Changes to the emitted X-ray spectra are negligible
  unless the thickness of mass leakage region at the base 
  approaches or exceeds one percent of the height of the post-shock region. 
In this case the predicted spectra from higher-mass white dwarfs will be harder, 
  and fits to X-ray data will predict lower white-dwarf masses 
  than previous formulations.
\end{abstract}


\section{Introduction}

In magnetic cataclysmic variables (mCVs) 
  the accreting material from the secondary star 
  is entrained onto magnetic field lines 
  and accretes onto the surface of the white dwarf. 
There it forms a standoff shock, followed by a hot post-shock region 
  of plasma settling onto the white dwarf surface 
  as it cools, principally by bremsstrahlung X-ray radiation, 
  and by optical/IR cyclotron radiation 
  if the magnetic field is sufficiently strong 
  (see Cropper 1990 and Warner 1995 for reviews of mCVs).
 
There have been a number of studies of the post-shock accretion flow
  (e.g.\ Aizu 1973; Langer, Chanmugam \& Shaviv 1981; 
  Chevalier \& Imamura 1982; Kylafis \& Lamb 1982; 
  Chanmugam, Langer \& Shaviv 1985; Imamura et\,al.\ 1987; 
  Wu 1994; Wu, Chanmugam \& Shaviv 1994; Woelk \& Beuermann 1996; 
  Kocabiyik 1997; Cropper et\,al.\ 1999; Saxton \& Wu 1999), 
  investigating different aspects of the region, 
  both analytically and numerically 
  (see Wu 2000 for a review). 
Despite the fact that substantial progress has been made, 
  the existing studies generally assume that 
  the velocity and the temperature drop to zero 
  at the base of the accretion column. 
The assumption of zero flow velocity 
  and the strict requirement of mass continuity along the field lines 
  immediately implies that the matter density 
  at the base of the accretion column must reach infinity. 
Recent work by Cropper, Wu \& Ramsay (2000) 
  has highlighted the consequences of an infinite density 
  at the base of the accretion column: 
  the X-rays from the shock-heated region is emitted mostly 
  from its base 
  where the density is rising steeply. 
This is true at energies up to approximately that of the shock temperature, 
  which is in the $\sim 10-60$~keV range ---  
  higher than most imaging or CCD-based X-ray instruments. 
The preponderance of emission from the base of the accretion column 
  has made a more in-depth understanding of the physical conditions 
  in shock-heated emission regions essential, 
not only from the theoretical point of view 
  but also for modelling and extracting information 
  from the observed X-ray spectra. 

Several aspects have to be considered 
  in order to derive a self-consistent formulation of the accretion flow, 
  especially at the boundary layer 
  where the hydrodynamic flow merges into a hydrostatic white-dwarf atmosphere. 
In the stationary-state case, the hydrodynamic variables 
  at the base of the accretion column 
  should match smoothly to the corresponding variables 
  at the white-dwarf atmosphere. 
In addition, the energy deposition due to the heat flux 
  emerging from below the white-dwarf atmosphere should be considered. 
Moreover, at some point in the post-shock region, 
  material may become sufficiently cold that 
  a fraction becomes neutral and cannot be efficiently confined 
  by the magnetic field. 
In the time-dependent situation, 
  it is necessary to consider the response of the white-dwarf atmosphere 
  to changes in the local mass-accretion rate and its effects 
  on the stability of the accretion flow. 
   
In this paper we investigate the stationary-state accretion 
  onto magnetic white dwarfs, 
  with emphasis on the post-shock flow at the boundary layer 
  above the white-dwarf atmosphere. 
We consider a modification to the conventional hydrodynamic formulation 
  so that matter diffusion across the field lines is allowed 
  in a thin region at the bottom of the accretion column. 
In addition, the heat flux from the white-dwarf atmosphere 
  into the accretion column is considered. 
As it is an exploratory study, we neglect the complications 
  such as geometrical, gravity and two-fluid effects. 
We demonstrate that a closed-form solution as that 
  described in Aizu (1973) and Wu et\,al.\ (1994) 
  (see also Chevalier \& Imamura 1982) can be obtained 
  for this formulation. 
Moreover, the infinities and discontinuities at the base, 
  which lead to unphysical observational consequences, are eliminated. 
We calculate the emission from the modified post-shock region and 
compare their spectral properties 
  with those of the conventional post-shock region. 

\section{Stationary-state accretion flow}

\subsection{Modified hydrodynamic formulation}

We assume that the flow of the ionised accreting material follows 
  the magnetic field lines 
  and that the field lines are perpendicular to the white-dwarf surface, 
  so the flow is (quasi-)one-dimensional. 
We neglect gravity effects, which are important 
  when the shock height $x_{\rm s}$ is not negligible 
  in comparison with the white-dwarf radius $R_{\rm w}$ 
  (see Cropper et\,al.\ 1999). 
We also assume that only the ionised matter 
  is strictly confined by the magnetic field, 
  while the cold neutral atomic matter 
  can ``leak'' out of the accretion column. 
The accretion matter obeys the ideal gas law, 
  i.e.\ $P = 2 \rho k T/m_{\rm p}$ and $\gamma = 5/3$, 
  where $P$ is the gas pressure, $T$ the gas temperature, 
  $k$ the Boltzmann constant, $m_{\rm p}$ the proton mass 
  and $\gamma$ the adiabatic index of the gas. 
At the lower boundary of the accretion column 
  the temperature gradient and the bulk-flow velocity are zero, 
  and the cooling rate equals the heating rate. 

In our formulation 
  the stationary state mass continuity, momentum 
  and energy equations are      
\begin{eqnarray}  
  {\partial \over {\partial  x}}{\rho v} & = & - \Sigma\ ,   \\ 
  {\partial \over {\partial  x}} P \ +\ 
    \rho v~{\partial \over {\partial  x}} v & = & v \Sigma \ ,  \\ 
   v~{\partial \over {\partial  x}} P \ +\ 
   \gamma  P~{\partial \over {\partial  x}} v  
          & = & - (\gamma -1)~\bigg[~\Lambda_{\rm c} - \Lambda_{\rm h} + 
             {1 \over 2}~v^2 \Sigma~\bigg]\ , 
\end{eqnarray}
  where $v$ is the flow velocity, and $\rho$ the density. 
$\Sigma$ is the sink term 
  specifying the rate of mass loss out from the accretion column; 
  $\Lambda_{\rm c}$ is the effective cooling function, 
  and $\Lambda_{\rm h}$ is the heating function. 
In the construction of the formulation, 
  we have implicitly assumed that the neutral matter, 
  which is not confined by the magnetic field, 
  transfers its momentum and energy to the ionised matter 
  before it leaves the accretion column. 
Under this assumption  
  momentum is conserved along each flow line of the ionised material and 
  energy is dissipated only by emitting radiation. 
In reality the neutral matter may carry away 
  substantial amount of momentum and energy, and 
  hence the strict condition for conservation of momentum and energy 
  along the field lines needs to be relaxed. 
This case will be discussed elsewhere (Wu \& Cropper, in preparation). 
  
The accretion shock is assumed to be strong and adiabatic, 
  with the immediate post-shock velocity equal 
  to a quarter of the pre-shock velocity, which 
  is taken to be the free-fall velocity at the white-dwarf surface,  
  i.e.\ $v = -v_{\rm ff}/4$ at $x = x_{\rm s}$. 
(Here and hereafter, the subscript ``s'' denotes variables at the shock.) 
The post-shock pressure is three-quarter 
  of the pre-shock ram pressure of the inflowing gas, 
  i.e.\ $P_{\rm s} = 3 \rho_{\rm a} v_{\rm ff}^2/4$, 
  where $\rho_{\rm a}$, the density of the pre-shock accretion matter, 
  is related to the specific accretion rate (per unit area) $\dot m$ 
  by $\rho_{\rm a} = \dot m /v_{\rm ff}$. 

We define the dimensionless variables $\xi = x/ x_{\rm s}$,  
  $\zeta = \rho / \rho_{\rm a}$, $\tau = - v/ v_{\rm ff}$ and 
  $\varpi = P/ \rho_{\rm a} v_{\rm ff}^2$, 
  and substitute them into the mass, momentum and energy equations. 
In terms of these variables, the hydrodynamic equations are  
\begin{eqnarray} 
 {\partial \over {\partial \xi}}{\zeta \tau} & = &  \tilde \Sigma \ ;  
     \\ 
 {\partial \over {\partial \xi}} \varpi\ +\ 
   \zeta \tau~{\partial \over {\partial \xi}} \tau  & = & 
   -\tau \tilde \Sigma \ ; \\ 
   \tau {\partial \over {\partial \xi}}\varpi \ +\ 
   \gamma \varpi ~{\partial \over {\partial \xi}} \tau  
   & = & (\gamma - 1)~\bigg[~\tilde \Lambda + 
    {1 \over 2}~\tau^2 \tilde \Sigma~\bigg]    \ ,
\end{eqnarray} 
  where $\tilde \Sigma = (x_{\rm s}/\rho_{\rm a} v_{\rm ff})\Sigma$, and 
$\tilde \Lambda =  (x_{\rm s}/\rho_{\rm a} v_{\rm ff}^3) 
  (\Lambda_{\rm c} - \Lambda_{\rm h})$.  
At the white-dwarf surface ($\xi = 0$), the velocity $\tau = 0$; 
  at the shock ($\xi = 1$), $\tau = 1/4$ and the pressure $\varpi = 3/4$.  

The sink term above, $\tilde \Sigma$, is not a direct observable, 
  but its integration over the distance along the flow 
  is the mass loss rate per unit area. 
It is therefore more appropriate 
  to consider the integration of the sink term over the distance 
  and express it in terms of the flow velocity, 
  the independent variable that we use in solving the hydrodynamic equations. 

In terms of the integration of the sink term over the distance 
  we can define a dimensionless variable    
\begin{equation} 
   \sigma (\tau) \ =\ 1 - \int^1_{\xi(\tau)} d\xi'~
    \tilde \Sigma (\xi') \ .  
\end{equation}
At the white-dwarf surface $\sigma (0) = 0$, 
  and at the shock $\sigma (1/4) = 1$.  
The variable $\sigma (\tau)$ clearly satisfies the mass continuity equation 
  and it is in fact the dimensionless specific mass accretion rate 
  in the presence of matter leakage. 

The dimensionless density can now be obtained 
  by solving the mass continuity equation, 
  and it is  
\begin{equation} 
  \zeta(\tau)\  =\  {{\sigma(\tau)} \over \tau}  \ .    
\end{equation} 
The dimensionless pressure, obtained by integrating the momentum equation, is  
\begin{equation} 
  \varpi(\tau) \ = \ 1 \  -  \ \tau~\sigma(\tau)  \ .  
\end{equation} 
Combining equations (8) and (9) with the energy equation yields 
\begin{equation}
 {\partial \over {\partial \xi}}\tau \ =\  
   (\gamma -1)~{\tilde \Lambda}~\bigg[~ 
  \gamma - (\gamma+1)~\tau \sigma -{1\over 2}~(\gamma+1)~\tau^2  
   {\partial \over {\partial \tau}} \sigma \bigg]^{-1}\ . 
\end{equation}
The above equation can be integrated 
  to obtain a closed-form expression for $\xi(\tau)$:  
\begin{equation} 
 \xi (\tau) \ =\ 1\ - \ \bigg({1 \over {\gamma -1}}\bigg) 
   \int^{1/4}_\tau d\tau'~{1 \over {\tilde \Lambda}}~ 
   \bigg[~\gamma~-(\gamma+1)~\tau' \sigma - {1\over 2}~(\gamma+1)~   
   \tau'^2 {\partial \over {\partial \tau'}} \sigma~ 
  \bigg] \ .   
\end{equation}  

We follow Wu (1994) and consider the cooling term $\Lambda_{\rm c}$ 
  (in equation [3]) in a composite-functional form  
\begin{equation}  
 \Lambda_{\rm c} \ =\ A \rho^2 \bigg({P \over \rho} \bigg)^{1/2} 
   \bigg[ 1 + \epsilon_{\rm s} \bigg({P \over {P_{\rm s}}} \bigg)^{\alpha} 
    \bigg({{\rho_{\rm s}} \over \rho} \bigg)^{\beta}\bigg]\ .    
 \end{equation}  
The parameter $\epsilon_{\rm s}$ specifies 
  the ratio of the bremsstrahlung cooling time-scale 
  to the cyclotron cooling time-scale at the shock. 
The value of the constant $A$ is $3.9 \times 10^{16}$ in c.g.s.\ units 
  for pure hydrogen plasmas (see Rybicki \& Lightman 1979), 
  and $\alpha \approx 2$ and $\beta \approx 3.85$ 
  for parameters appropriate for mCVs. 
The dimensionless cooling/heating term $\tilde \Lambda$ is then  
\begin{equation} 
 \tilde \Lambda \ = \  
   \bigg({{x_{\rm s}}\over {\rho_{\rm a} v_{\rm ff}^3}} \bigg)~ 
   \bigg\{~ \rho_{\rm a}^2 v_{\rm ff}A~ 
   \bigg({\sigma \over \tau} \bigg)^{3/2} \sqrt{1-\tau\sigma}~ 
    \bigg[~1\ +\ \epsilon_{\rm s} {{4^{\alpha + \beta}} 
    \over {3^{\alpha}}}~ 
     \bigg(1-\tau \sigma \bigg)^{\alpha} 
   \bigg({\tau \over \sigma} \bigg)^{\beta} \bigg] 
   \ - \ \Lambda_{\rm h}(\tau) \bigg\}\ .   
\end{equation}   
With the cooling function is defined, 
  the flow velocity, and hence the other hydrodynamic variables, can be derived 
  after equation (11) is inverted.  

If we set $\gamma = 5/3$, $\sigma (\tau) = 1$ and $\Lambda_{\rm h} = 0$, 
  equation (11) becomes  
\begin{equation} 
 \xi(\tau) \ = \  
    {{v_{\rm ff}^2} \over {2 A x_{\rm s} \rho_{\rm a}}} \int_0^{\tau} 
    d\tau'~ 
  {{\tau'^2~(5-8\tau')}\over {\sqrt{\tau'(1-\tau')}}} 
  \bigg[~1\ +\ \epsilon_{\rm s} {{4^{\alpha + \beta}} 
  \over {3^{\alpha}}}~ 
     (1-\tau')^{\alpha} \tau'^{\beta} \bigg]^{-1}\ , 
\end{equation} 
  which is identical to equation (3) in Wu (1994). 
This limiting case corresponds to ``no leaking'' from the accretion column 
  and a hard, stationary white-dwarf surface 
  with no emergent energy flux from below. 
If we further set $\epsilon_{\rm s} = 0$, the Aizu (1973) solution is recovered.  

\subsection{The sink term} 
 
When the accreting material is cooled sufficiently, 
  recombination occurs and neutral atoms are formed. 
While the charged ionised matter is magnetically confined, 
  neutral matter can diffuse across the field lines. 
As the temperature is lower near the base of the accretion column, 
  we expect matter leakage to be efficient.  

In the hydrodynamic formulation that we consider, 
  the temperature and the flow velocity are not independent. 
The temperature is a monotonic function of the velocity, 
  and in the limit of no leakage 
  it is linearly proportional to the velocity 
  at the base of the accretion column. 
The sink (which depends on the hydrodynamic variables) 
  can therefore be chosen to be an explicit function of the flow velocity only. 
We assume that it is characterised by a critical velocity $\tau_{\rm m}$, 
  below which matter leakage becomes efficient. 
This is equivalent to assuming a critical temperature 
  at which recombination becomes sufficiently efficient 
  to allow significant amount of neutral matter to form 
  and to diffuse out of the accretion column. 

\begin{figure}
\vspace*{0.4cm} 
\begin{center} 
 \epsfxsize=8.5cm 
 \epsfbox{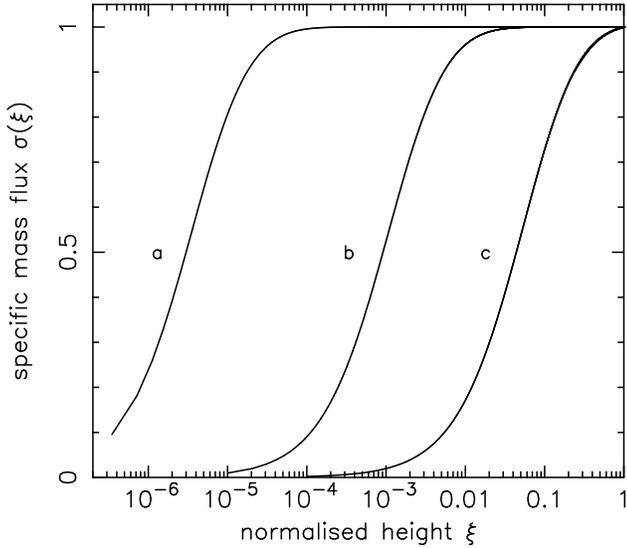} 
\end{center}
\caption{ 
Specific mass flux $\sigma(\xi)$ (mass transfer rate per unit area) 
  as a function of height $\xi$ within the post-shock region 
  for $\tau_{\rm m}$ = 0.001, 0.01 and 0.05 
  (curves a, b and c respectively). }
\end{figure} 

It is more convenient to consider $\sigma(\tau)$ 
  (the specific mass-accretion rate per unit area 
  remaining in the post-shock flow at height $\xi$ after leakage) 
  than the leakage at any height $\tilde \Sigma(\xi)$ itself. 
Suppose $\sigma(\tau)$ can be expanded into a series of orthogonal bases  
  $e^{-n\tau/\tau_{\rm m}}$, 
  such that $\sigma(\tau) = \sum a_n e^{-n\tau/\tau_{\rm m}}$, 
  where $a_n$ is the coefficient of the $n$-th power term. 
If only the first two leading terms are important, 
  as an approximation we may consider 
\begin{equation} 
  \sigma(\tau)\ =\ \mu~\bigl( 1 - \nu e^{-\tau/\tau_{\rm m}} \bigr), 
\end{equation}  
  where $\mu$ and $\nu$ are constants to be determined. 
In this study we treat $\tau_{\rm m}$ as a parameter. 
When appropriate atomic and magneto-hydrodynamic processes are considered, 
  an explicit expression of it in terms of the other system parameters 
  can be obtained. 

The functional form of $\sigma(\tau)$ above 
  must satisfy the hydrostatic-equilibrium condition 
  at the white-dwarf surface ($\xi = 0$). 
This requires the mass flux $\sigma(\tau)$ equal to zero 
  at the white-dwarf surface. 
Thus, the constant $\nu = 1$. 
At the shock ($\xi =1$), the mass flux $\sigma(\tau) = 1$ 
  by definition (equation 7), 
  implying a normalisation constant 
  $\mu = (1 - e^{-1/4\tau_{\rm m}})^{-1}$. 
In the limit of $\tau_{\rm m} \ll 1$, 
  the velocity-derivative of $\sigma(\tau)$ is
\begin{equation}  
  {\partial \over {\partial \tau}} \sigma \approx {1\over \tau_{\rm m}} 
\end{equation}   
  at the white-dwarf surface, and it is  
\begin{equation}  
 {\partial \over {\partial \tau}} \sigma  \approx 0      
\end{equation} 
  at the shock. 
Matter leakage therefore occurs 
  only at the very bottom of the accretion column 
  if $\tau_{\rm m}$ is sufficiently small, 
  and mass is practically conserved along the flow 
  in most of the post-shock region. 
For $\tau_{\rm m} = 0.01$, the specific mass flux 
  falls to less than 1/2 of its initial value 
  only when the normalised height $\xi\ \LS \ 0.001$; 
  and for $\tau_{\rm m} = 0.001$, 
  when $\xi\ \LS \ 3\times 10^{-6}$ (Fig.~1).  

\subsection{Energy flux from the white-dwarf atmosphere}   

We assume that the heating due to the energy flux 
  from below the white-dwarf surface is important 
  only in a thin region (with a thickness $\Delta \tau < \tau_{\rm m}$) 
  at the bottom of the accretion column. 
Suppose $\pi S_{\rm w} (0)$ is the radiation flux 
  emerging normally from the white-dwarf surface. 
The radiation is absorbed by the accreting matter 
  and is attenuated. At the height $\xi$, 
  the radiation flux is given by 
\begin{equation} 
 \pi S_{\rm w}(\xi) \ = \ 
  \pi S_{\rm w}(0)~\exp~\biggl(-~x_{\rm s} \int _0^{\xi} 
  d\xi'~\kappa(\xi')\biggr) \,  
\end{equation}
  where $\kappa(\xi)$ is the absorption coefficient. 
The rate of energy deposited per unit volume is therefore   
\begin{equation} 
 -{{\pi} \over x_{\rm s}} {d \over {d\xi}} S_{\rm w}(\xi) 
 \ = \ \pi~\kappa(\xi) 
 S_{\rm w}(0)~\exp~\biggl(-~x_{\rm s} 
 \int _0^{\xi} d\xi'~\kappa(\xi')\biggr)\ . 
\end{equation}   

The absorption coefficient $\kappa$ 
  depends only on the local density and temperature, 
  which are approximately constant when $\tau < \tau_{\rm m}$ 
  (see next section). 
In local thermal equilibrium, 
  the absorption coefficient is determined by the emissivity, 
  which is the effective cooling function in our formulation, 
  and the Planck function ${\cal B}$ (Kirchoff's Law). 
As the emergent flux from the white dwarf is a black-body flux, 
  i.e.\ $S_{\rm w}(0) = {\cal B}(0)$, we have 
\begin{eqnarray}  
  \kappa(\xi) S_{\rm w}(0) & \approx & \kappa(0) S_{\rm w}(0) \nonumber \\ 
   & = & {1 \over \pi} \Lambda_{\rm c}(0) \ .   
\end{eqnarray} 
Moreover, 
\begin{eqnarray} 
 \int_0^{\xi} d\xi'~\kappa(\xi') & = &  
 \int_0^{\xi} d\xi'~ \biggl[ {{\Lambda_{\rm c}(\xi') }  
  \over {{\cal B}(\xi')}} \biggr] \nonumber    \\  
   & \approx  & {{\xi~\Lambda_{\rm c}(0)}  
  \over {{\cal B}(0)}}  \ .  
\end{eqnarray}   
 
We have defined the lower boundary of the accretion column 
  as the location where the heating rate equals the cooling rate 
  and the flow velocity is zero. 
This requires $\Lambda_{\rm h} (\tau) = \Lambda_{\rm c} (\tau)$ 
  and $\tau = 0$ at $\xi = 0$. 
Suppose the attenuation factor can be expressed 
  in terms of the dimensionless velocity $\tau$, such that  
\begin{equation} 
  x_{\rm s} \int_0^{\xi(\tau)} d\xi'~\kappa(\xi')  
       = {\sqrt{\tau} \over \lambda} + {\cal O}(\tau) \ , 
\end{equation} 
  where $\lambda$ is a constant to be determined. 
Then, the heating function in the boundary layer is    
\begin{equation} 
 \Lambda_{\rm h}(\tau) \approx 
  \Lambda_{\rm c}(0)~\exp~\bigg(-{\sqrt{\tau} \over \lambda}\bigg) \ . 
\end{equation}  

By expanding $\tau$ into Taylor series at the boundary $\xi_{\rm o} = 0$, 
  we obtain   
\begin{equation} 
\sqrt{\tau}  \ = \ 
  \biggl[ \xi~  
   {{\partial \tau} \over {\partial \xi}}_{\xi_{\rm o}=0}   
   +\ {{\xi^2}\over {2!}}
   {{\partial^2 \tau} \over {\partial \xi^2}}_{\xi_{\rm o}=0} 
   +\ ........ \   
   \biggr]^{1/2} \ . 
\end{equation}   
Smooth merging of the flow into a hydrostatic white-dwarf atmosphere 
  requires the first derivative of the velocity to be zero. 
By combining equations (10), (23) and (24)   
  and keeping only the first non-vanishing term, we obtain  
\begin{equation}  
  {\sqrt \tau}\ = \   
  {\xi \over {2 \lambda}} \biggl( {{\gamma -1} \over \gamma} \biggr) 
     \biggl({{x_{\rm s} \Lambda_{\rm c}(0)} 
   \over {\rho_{\rm a} v_{\rm ff}^3}} \biggr)   \ . 
\end{equation}  

We can now define a characteristic critical velocity $\tau_{\rm h}$, 
  below which heating by the flux 
  from below the white-dwarf atmosphere is important 
  and beyond which heating effects are negligible.  
$\tau_{\rm h}$ is given by 
\begin{eqnarray} 
 \tau_{\rm h} & \equiv & \lambda^2 \nonumber \\ 
   & = &  {1 \over 2} \biggl( {{\gamma -1} \over \gamma} \biggr) 
  \biggl({{{\cal B}(0)}\over{\rho_{\rm a} v_{\rm ff}^3}} \biggr) \  .   
\end{eqnarray} 
The heating function 
\begin{equation} 
  \Lambda_{\rm h}(\tau) \ = \  
  \Lambda_{\rm c}(0)~\exp~\bigg(-{\sqrt{{\tau}\over{\tau_{\rm h}}}}\bigg)   
\end{equation}   
  is asymptotically zero for $\tau \gg \tau_{\rm h}$. 
As heating is unimportant for $\tau > \tau_{\rm m} > \tau_{\rm h}$, 
  we can simply use the heating function above 
  throughout the whole post-shock region, 
  despite the fact it is derived 
  from considering heating of the accreting material 
  within the boundary layer.    

\section{Properties of the Hydrodynamic Variables at the Lower Boundary}

\subsection{Hydrodynamic variables} 

The matter density at the base of the ``leaky'' accretion column 
  that we consider is 
\begin{eqnarray} 
 \zeta (0) & = & {\lim_{\tau \to 0}}~ {\mu \over \tau}~  
   \bigl( 1 - e^{-\tau/\tau_{\rm m}} \bigr) \nonumber \\ 
  & = & {\mu \over \tau_{\rm m}} \ ,       
\end{eqnarray}
  which is clearly finite. 
The pressure is 
\begin{eqnarray} 
 \varpi (0) & = & {\lim_{\tau \to 0}}~ \big[ 1- {\tau \mu }~  
   \bigl( 1 - e^{-\tau/\tau_{\rm m}} \bigr) \big] \nonumber \\ 
  & = & 1 \ ,    
\end{eqnarray} 
  the same as that in the conventional formulation. 
The dimensionless temperature is defined as $\theta \equiv {T/ T_{\rm s}}$, 
  and hence we have 
\begin{equation} 
  \theta \ =\ {\zeta_{\rm s} \over \varpi_{\rm s}} 
   {\varpi \over \zeta}\ .  
\end{equation}  
At the white-dwarf surface, the temperature is   
\begin{eqnarray} 
   \theta (0) & = & 
   {\lim_{\tau \to 0}}~ {16 \over 3} {\tau \over \mu}
  \bigg[{{1- \tau \mu ( 1 - e^{-\tau/\tau_{\rm m}} \bigr)} 
  \over { 1 - e^{-\tau/\tau_{\rm m}} }} \bigg] \nonumber \\ 
  & = & {16 \over 3}{\tau_{\rm m} \over \mu} \ ,     
\end{eqnarray} 
  which is also finite. 

In the limit of $\tau_{\rm m} \rightarrow 0$, 
  we have $\sigma(\tau) \rightarrow 1$ and 
  $\partial \sigma/\partial \tau \rightarrow 0$. 
Moreover, $\zeta(0) \rightarrow \infty$, 
  $\theta(0) \rightarrow 0$, and $\varpi(0) = 1$, 
  i.e.\ the cold ``stationary-wall'' boundary condition is recovered.   

\subsection{Gradients of the hydrodynamic variables}  

Not only are the hydrodynamic variables of the conventional 
  and our formulations are different at the lower boundary, 
  but they also have very different gradients. 
In the conventional formulation, the density gradient is given by 
\begin{equation}  
 {\partial \over {\partial \xi}} \zeta \ =\ 
  - {1\over {\tau^2}}{\partial \over {\partial \xi}} \tau \ .  
\end{equation} 
As $\partial \tau/ \partial \xi \propto \tau^{-3/2}$ for small $\tau$ 
  (from equation (2) in Wu 1994), 
  the density gradient is proportional to $\tau^{-7/2}$. 
At $\xi = 0$, $\tau = 0$; hence the density gradient is infinite. 
The pressure gradient, which is 
\begin{eqnarray}  
 {\partial \over {\partial \xi}} \varpi & = &  
  - {\partial \over {\partial \xi}} \tau \nonumber \\ 
  & \propto &    - \tau^{-3/2} \  ,   
\end{eqnarray} 
  and the temperature gradient, which is 
\begin{eqnarray}  
 {\partial \over {\partial \xi}} \theta & = &  
  {16 \over 3} (1 - 2 \tau)\ 
  {\partial \over {\partial \xi}} \tau \nonumber \\ 
  & \propto &    - \tau^{-3/2} \  ,   
\end{eqnarray} 
  are also infinite at $\xi = 0$; 

For a ``leaky'' accretion column 
  allowing an energy flux emerging from below, 
  the velocity gradient at the base is 
\begin{eqnarray}
 {\lim_{\tau \to 0}}~{\partial \over {\partial \xi}}\tau & = &   
  {\lim_{\tau \to 0}}~ (\gamma -1)~{\tilde \Lambda}~\bigg[~ 
  \gamma - (\gamma+1)~\tau \sigma -{1\over 2}~(\gamma+1)~\tau^2  
   {\partial \over {\partial \tau}} \sigma \bigg]^{-1}\  \nonumber \\ 
 &  = & \biggl[
  {{(\gamma -1)x_{\rm s}}\over{\gamma \rho_{\rm a} v^3_{\rm ff}}}\biggr]~
  (\Lambda_{\rm c}(0)-\Lambda_{\rm h}(0)) \ .  
\end{eqnarray}  
The velocity gradient is zero at $\xi = 0$, 
  provided that $\Lambda_{\rm c}(0) = \Lambda_{\rm h}(0)$. 
This is in fact the condition of radiative equilibrium 
  in a hydrostatic white-dwarf atmosphere. 

Differentiating equation (8) with respect to $\xi$ 
  yields the density gradient:    
\begin{equation} 
 {\partial \over {\partial \xi}} \zeta \  = \ - {1 \over \tau} 
  \bigg[ {\sigma \over \tau} - 
  {{\partial \over {\partial \tau}} \sigma} \bigg]  
  {\partial \over {\partial \xi}} \tau \ . 
\end{equation}  
By substituting the expression of $\sigma(\tau)$ given in equation (15) 
  into the equation above, we obtain      
\begin{equation}  
 {\partial \over {\partial \xi}} \zeta \  
 \ = \   {\mu \over {\tau_{\rm m}^2}}   
  \biggl[ \sum_{n=0}^{\infty} 
  C_n \bigg({\tau \over \tau_{\rm m}} \bigg)^n \biggr]
    {\partial \over {\partial \xi}} \tau 
  \  \ \ \   (\tau < \tau_{\rm m})\ ,    
\end{equation}   
  where 
\begin{equation} 
  C_n\ =\ (-1)^n \bigg[  {1\over {(n+1)!}} 
   - {1 \over {(n+2)!}}  \bigg]\ .    
\end{equation} 
The power series $C_n$ converges, with a value of the order of unity 
  when $\tau \rightarrow 0$ (Appendix A.1). 
The density gradient is therefore finite and equals zero at $\xi = 0$.   

The pressure gradient is 
\begin{equation} 
 {\partial \over {\partial \xi}} \varpi \  = \  -\mu 
  \biggl[ 1- 
  \biggl( 1- {\tau \over \tau_{\rm m}} \biggr) e^{-\tau/\tau_{\rm m}} 
   \biggr]   {\partial \over {\partial \xi}} \tau \ .   
\end{equation}  
It is also finite and equals zero at $\xi = 0$. 
The temperature gradient is 
\begin{equation} 
 {\partial \over {\partial \xi}} \theta \  =  \   
 {{\zeta_{\rm s}} \over {\varpi_{\rm s}}} 
 \biggl[ {1\over \zeta^2} 
 \biggl(\zeta {\partial \over {\partial \tau}} \varpi - 
   \varpi {\partial \over {\partial \tau}} \zeta \biggr) \biggr] 
   {\partial \over {\partial \xi}} \tau \ . 
\end{equation}   
As $\partial \varpi/ \partial \tau$ and $\partial \zeta / \partial \tau$ 
  are both finite and the density $\zeta$ is non-zero at $\xi = 0$, 
  the temperature gradient is zero.  

\section{Accretion Luminosity} 

In the conventional formulation, energy conservation holds 
  along the field (flow) line, 
  so that the accretion energy must be radiated away 
  before the accreting matter can settle down onto the white-dwarf surface. 
The accretion luminosity (normalised to $\rho_{\rm a} v_{\rm ff}^2$) 
  along the flow line down to the height $\xi$ is    
\begin{eqnarray} 
  L_{\rm acc}(\xi,1) & = &  
     \int_{\xi}^1 d\xi \ {\tilde \Lambda} \nonumber \\ 
      & = & {1\over {\gamma - 1}} \int_{\tau}^{1/4} d\tau\ 
       [\gamma -(\gamma+1)\tau ] \nonumber  \\ 
      & = & {1\over {\gamma - 1}} \bigg\{ \biggl[ 
       {\gamma \over 4} - {1\over 32} (\gamma+1)  \biggr]  
       - \gamma \tau \biggl[  1 - \biggl(
      {{\gamma +1} \over {2 \gamma}} \biggr) \tau  \biggr] \bigg\} \ .  
\end{eqnarray}  
The total bolometric accretion luminosity is therefore 
\begin{equation} 
  L_{\rm acc}(0,1) \ = \  {1\over {\gamma - 1}} \biggl[ 
       {\gamma \over 4} - {1\over 32} (\gamma+1)  \biggr] \ .  
\end{equation}  
For $\gamma =5/3$, $L_{\rm acc} (0,1) = 1/2$, 
  consistent with the fact 
  that the total accretion energy liberated 
  is independent of the nature of the cooling process.  

For the ``leaky'' accretion column considered here, energy conservation 
  also holds strictly along the flow lines. 
The bolometric accretion luminosity along a field line down to $\xi$ is 
\begin{eqnarray} 
  L_{\rm acc}(\xi,1) & = & {1\over {\gamma - 1}} \int_{\tau}^{1/4} d\tau\ 
       \biggl[\gamma -(\gamma+1)\tau\sigma -{1\over 2}(\gamma+1)\tau^2 
        {{d\sigma}\over {d\tau}} \biggr]  \nonumber \\ 
      & = & {1\over {\gamma - 1}} \bigg\{ \biggl[ 
       {\gamma \over 4} - {1\over 32} (\gamma+1)  \biggr]  
     -   \gamma \tau \biggl[  1- \biggl(
      {{\gamma +1} \over {2 \gamma}} \biggr) \biggl(
      {{1 - e^{-\tau/\tau_{\rm m}}} 
     \over {1-e^{-1/4\tau_{\rm m}}}} \biggr) \tau 
       \biggr] \bigg\} \    
\end{eqnarray}   
  (Appendix A.2), where $\sigma$ is as defined in equation (15), 
  and the total bolometric accretion luminosity is 
\begin{equation} 
  L_{\rm acc}(0,1) \ = \ {1\over {\gamma - 1}} \biggl[ 
       {\gamma \over 4} - {1\over 32} (\gamma+1)  \biggr]   \  .       
\end{equation}   
For $\gamma = 5/3$ it is equal to 1/2, 
  consistent with the requirement of strict energy conservation 
  along a field line. 

The bolometric accretion luminosity of the boundary layer, 
  where $\tau < \tau_{\rm m}$, is  
\begin{eqnarray}   
  L_{\rm acc}(0,\xi_{\rm m}) &  = &  
    L_{\rm acc}(0,1) - L_{\rm acc}(\xi_{\rm m},1) \nonumber \\ 
    & = & \biggl({{\gamma~\tau_{\rm m}} \over {\gamma - 1}}\biggr) 
   \biggl[ 1 - {{\gamma +1} \over {2 \gamma}} \biggl( 
   {{1-e^{-1}} \over {1- e^{-1/4\tau_{\rm m}}}} \biggr)
   \tau_{\rm m} \biggr]    \ , 
\end{eqnarray}  
  where $\xi_{\rm m}$ is the height at which $\tau = \tau_{\rm m}$. 
If the boundary layer is thin (i.e.\ $\tau_{\rm m} \ll 1$), 
  the bolometric luminosity $\sim \gamma \tau_{\rm m}$, 
  which is insignificant in comparison to 
  that of the rest of the post-shock region.   

\begin{figure}
\vspace*{0.4cm} 
\begin{center} 
 \epsfxsize=8.5cm 
 \epsfbox{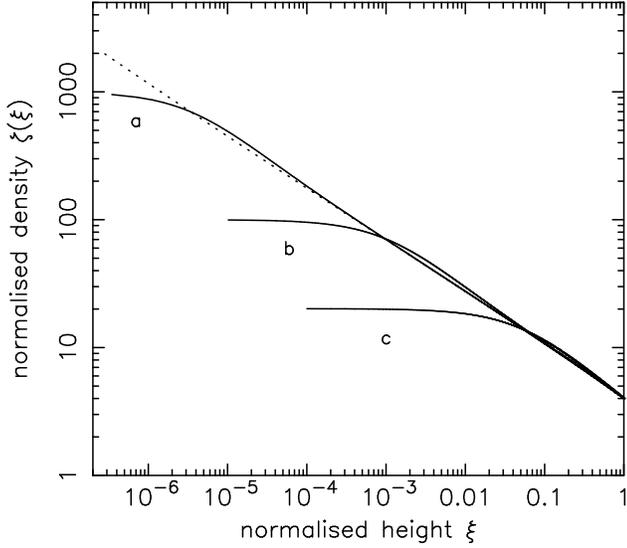} 
\end{center}
\caption{ 
Density $\zeta (\xi)$ as a function of height $\xi$ 
  for $\tau_{\rm m}$ = 0.001, 0.01 and 0.05 
  (curves a, b and c respectively). 
The density dependence of the conventional model (Aizu 1973) 
  is also shown for comparison (dotted line). }
\end{figure} 

\section{Discussion}

\subsection{Structure of the post-shock region}

Figure~2 shows the density profile at the base of the post-shock region 
  for ``leaky'' accretion columns with different values of $\tau_{\rm m}$ 
  by comparison with the conventional formulation (Aizu 1973), 
  in which the lower boundary is a cold, stationary wall. 
As noted above, the density $\zeta$ and its gradient 
  in the conventional formulation approach infinity 
  at the base of the accretion column, 
  whereas the density in our formulation has a finite value, 
  $\mu/\tau_{\rm m}$. 
For $\tau_{\rm m} = 0.05$, 0.01 and 0.001, 
  $\zeta(0) \approx 20$, 100 and 1000 respectively. 
In all cases of $\tau_{\rm m}$, 
  the density at the ``leaky'' base falls significantly 
  below that of the base in the conventional formulation 
  for $\xi$ less than  $\xi_{\rm m}$, 
  the height at which $\tau = \tau_{\rm m}$.  

\begin{figure}
\vspace*{0.4cm}   
\begin{center}  
\begin{tabular}{cc} 
 \epsfxsize=8.5cm 
 \epsfbox{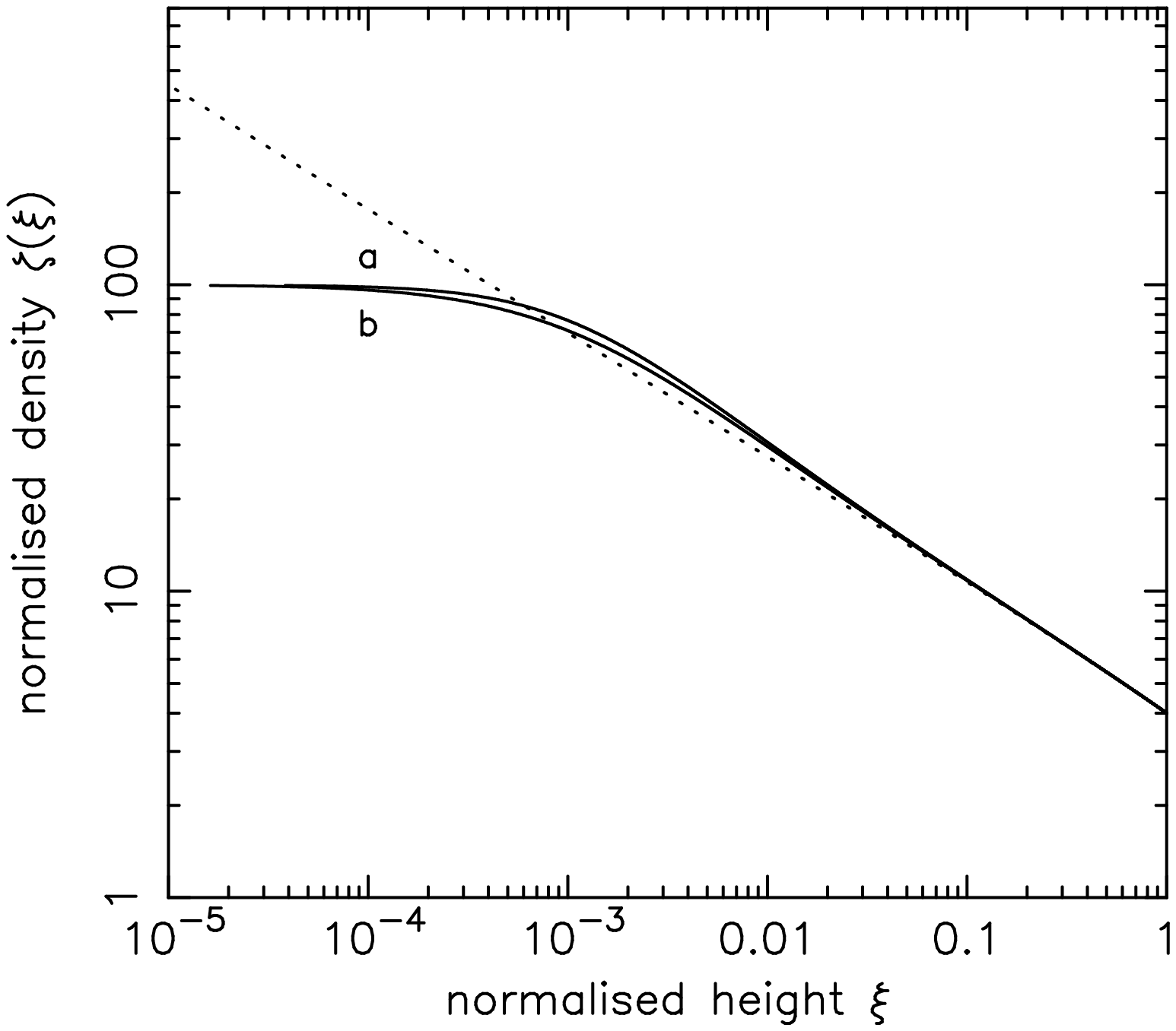} & 
 \epsfxsize=8.5cm 
 \epsfbox{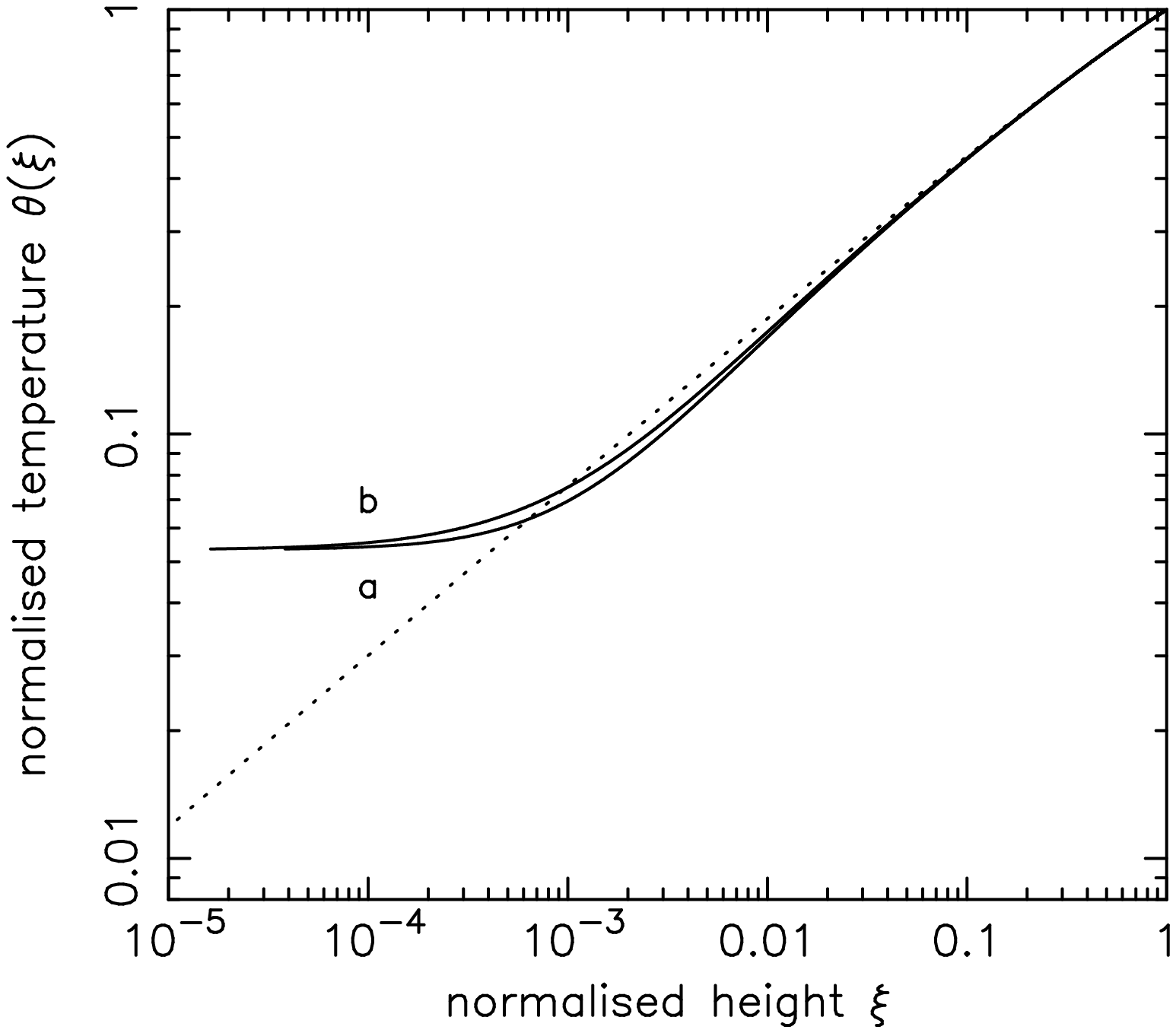}  
\end{tabular} 
\end{center}
\caption{ 
Density (left) and temperature (right) 
  as a function of height $\xi$ for heating scale height 
  $\tau_{\rm h}$ = 0.001 and 0.0001 (curves a and b respectively). 
The leakage parameter $\tau_{\rm m}$ is fixed to be 0.01. 
The conventional model is shown (in dotted line) for comparision. }
\end{figure} 

In calculating the density profiles in Figure~2, 
  we have not included the effect due to heating from the white dwarf, 
  i.e.\ we have considered $\tau_{\rm h}=0$. 
We now show this effect in Figure 3, 
  where we have fixed the critical velocity $\tau_{\rm m}$ to 0.01, 
  and calculated the dimensionless density and temperature 
  for two different boundary layers, 
  with $\tau_{\rm h} = 0.001$ and $0.0001$. 
They both fulfill the constraint that $\tau_{\rm h} < \tau_{\rm m}$. 
As shown, the terminated temperature depends mainly on $\tau_{\rm m}$. 
The heat input from the white dwarf modifies 
  only the asymptotic properties of the base temperature 
  --- a larger $\tau_{\rm h}$ will result in 
  a slightly thicker ``isothermal'' layer at the base, 
  thus reducing the temperature upstream 
  and causing the density to reach the terminated plateau value 
  further upstream. 

Figure~4 shows the dimensionless local bremsstrahlung emissivity 
  $j_{\rm br}$ ($= \zeta^2 \theta^{1/2}$) 
  as a function of the height $\xi$ in the post-shock region. 
The emissivity reaches a terminated value determined by $\tau_{\rm m}$, 
  whereas that from the conventional formulation tends to infinity. 
Again the bremsstrahlung emissivity is modified slightly 
  by the input of heat from the white dwarf 
  when $\tau_{\rm h} > 0$ 
  with increased emission for larger values of $\tau_{\rm h}$. 
This is in fact the consequence of an implicit assumption 
  that the white-dwarf energy flux is much weaker than energy flux 
  released by accretion material.    

\begin{figure}
\vspace*{0.4cm} 
\begin{center}  
 \epsfxsize=8.5cm 
 \epsfbox{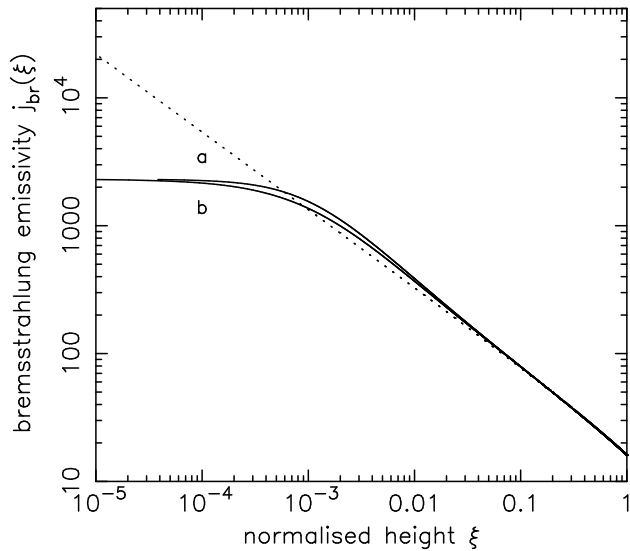}  
\end{center}
\caption{ 
Bolometric bremsstrahlung emissivity 
  $j_{\rm br}\ (=\ \zeta^2 \theta^{1/2})$ 
  as a function of height $\xi$. 
Curve a corresponds to the case of 
  $\tau_{\rm m}$  = 0.01 and $\tau_{\rm h}$ = 0.001; 
  curve b to the case of $\tau_{\rm m}$ = 0.01 and $\tau_{\rm h}$ = 0.   
The bremsstrahlung emissivity of the conventional model 
  is shown (in dotted line) for comparison. }
\end{figure} 

\subsection{Optical depths}

Using the prescription for density and temperature above 
  we can easily show that the optical depth 
  due to free-free absorption or electron scattering 
  in the vertical upstream direction is negligible 
  when the photon energies are above 0.05~keV, 
  for typical mass-transfer rates and white-dwarf masses. 
However, by assigning the parameter $\tau_{\rm h}$ 
  for the heating at the base of the region 
  we implicitly assume a large opacity 
  to maintain a black-body spectrum for the radiative flux. 
As the temperature in the region of our interest 
  should be sufficiently low 
  such that recombination can occur to form neutral atoms, 
  there are additional sources of opacity, 
  such as the bound-free and bound-bound opacities, 
  to maintain a large optical depth 
  in this geometrically thin layer at the base.  

The bound-free and bound-bound emission processes 
  are, however, not considered explicitly 
  in constructing the heating and cooling functions in our formulation. 
Instead, we consider only a parametric heating function $\Lambda_{\rm h}$, 
  which is in terms of the accretion parameters 
  and the energy flux from below the white-dwarf surface. 
The formulation that we present in this work 
  is therefore not a fully self-consistent formulation 
  with explicitly consideration of the microscopic atomic physics 
  (see van Teeseling, Heise \& Paerels 1994).  
  
Also, we have not considered an explicit treatment 
  of the ionisation equilibrium. 
There might be situations 
  that the gas at the base is photo-ionised by the X-ray from the shock above. 
The increase in the degree of ionisation 
  will cause an decrease in the efficiency of matter diffusion 
  across the field lines, 
  and hence out of the accretion column to the white-dwarf atmosphere. 
For typical mCV parameters, 
  say white-dwarf mass $M_{\rm wd} = 0.7$~M$_\odot$, 
  specific mass-accretion rate $\dot m \sim$1~g~cm$^{-2}$s$^{-1}$,  
  X-ray luminosity $L_{\rm x}$ of $\sim 10^{32}$~erg~s$^{-1}$, 
  shock temperature $T_{\rm s} \sim$10~keV 
  and shock height $x_{\rm s} \sim 3 \times 10^7$~cm, 
  the ionisation parameter $\Xi$   
  ($\equiv L_{\rm x}/n r^2$, 
  where $n$ is the number density 
  and $r$ is the distance to the X-ray source)   
  has values $\LS 0.01$.  
(Here, we have assumed $\tau_{\rm m} \approx 5.0 \times 10^{-5}$ 
  in the calculation of $\Xi$.)
The value of $\Xi$ is significantly less than 20, 
  the critical value at which hydrogen becomes completely ioinised 
  (see Models 3 and 4 in Kallman \& McCray 1982). 
Hence, the formulation presented above  
  is in general applicable to the accretion column of mCVs. 
Such a parametric prescription allows 
  to derive a analytic formulation to describe the accretion column 
  and the boundary layer in a simple and unified manner.  

It is worth noting that the surface 
  at which the total scattering and absorption optical depth is unity 
  is in effect the observable lower boundary of the post-shock region. 
This surface does not necessarily coincide with the base 
  as that defined by the flow hydrodynamics, the zero-velocity surface. 
The emission from this unit-optical-depth surface will be observed 
  as soft X-rays with a blackbody-like spectrum 
  characterised by a temperature similar to 
  that of the isothermal layer (due to $\tau_{\rm h} < \tau_{\rm m}$). 

\subsection{X-ray spectral properties}

With the density and temperature specified within the post-shock region 
  we can calculate the emission at different heights 
  and construct integrated spectra. 
We consider the case in which bremsstrahlung radiation 
  is the only cooling process and ignore cyclotron cooling, 
  so to illustrate the differences between between it 
  and the conventional post-shock X-ray emission regions 
  resulting from the Aizu (1973) formulation.  

\begin{figure}
\vspace*{0.4cm} 
\begin{center} 
\begin{tabular}{cc} 
 \epsfxsize=8.5cm 
 \epsfbox{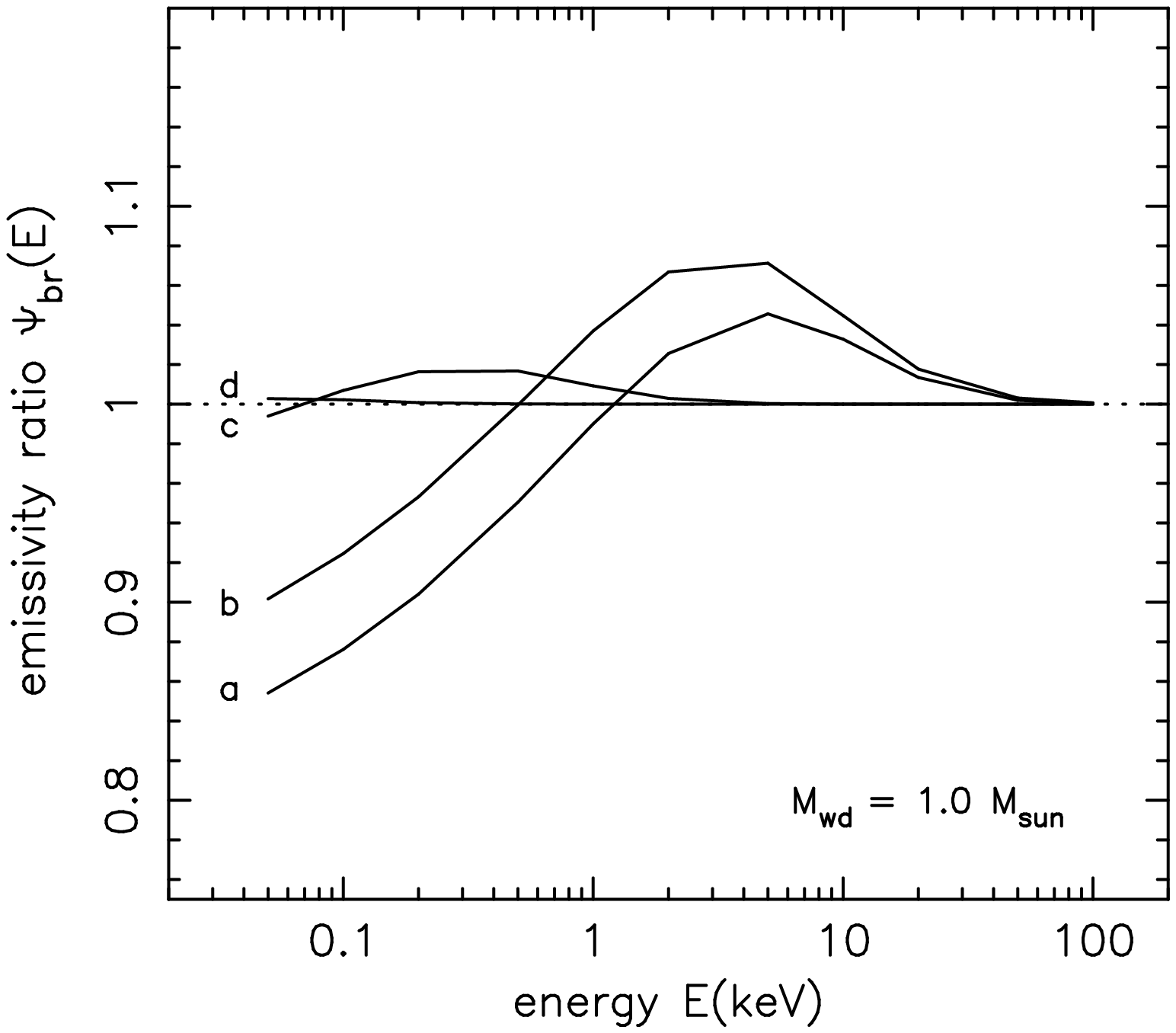} & 
 \epsfxsize=8.5cm  
 \epsfbox{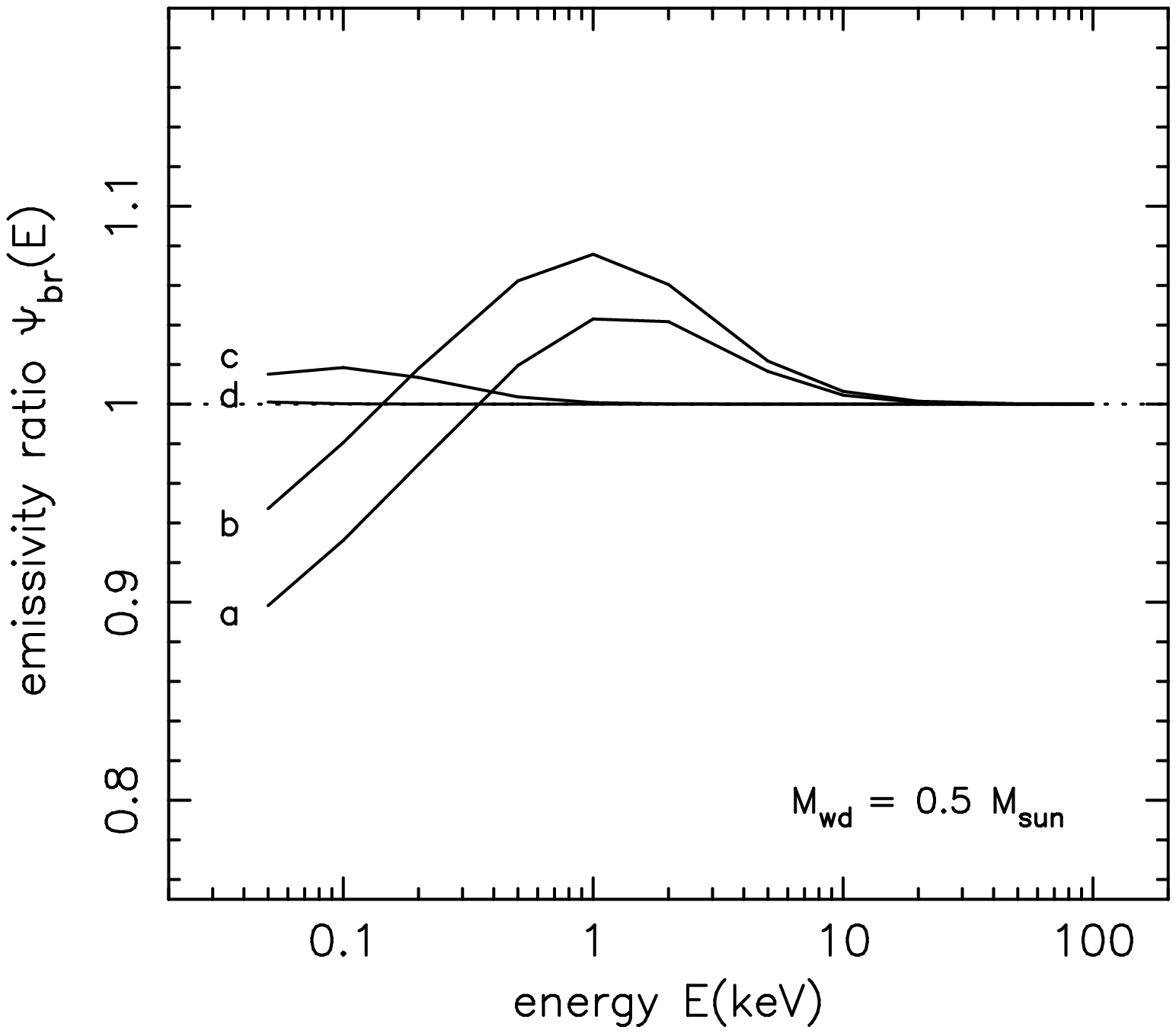} 
\end{tabular}
\end{center}
\caption{ 
The ratio of total bremstrahlung emissivity 
  of the ``leaky'' accretion column 
  to that of the conventional accretion column $\psi_{\rm br}$ 
  as a function of photon energy $E$ 
  for a 1.0-$\Msun$ white dwarf (left) and a 0.5-$\Msun$ white dwarf (right). 
The leakage and heating parameters are: 
  $(\tau_{\rm m},\tau_{\rm h})$ 
  = (0.01,0), (0.01,0.001), (0.001,0.0001) and (0.0001,0.00001) 
  for cuves a, b, c and d respectively. 
The specific mass flux is 1 g~s$^{-1}$cm$^{-2}$ at the shock ($\xi = 1$). }
\end{figure} 

Figure~5 illustrates the differences in the emitted spectrum 
  with respect to the conventional formulation 
  for 1.0-$\Msun$ and 0.5-$\Msun$ white dwarfs. 
The variable $\psi_{\rm bf}(E)$ 
  is the ratio of the bremsstrahlung emissivity 
  for the ``leaky'' post-shock region $j_{\rm br}(E;\tau_{\rm m})$ 
  to that for the conventional region $j_{\rm br}(E;0)$, 
  where $E$ is the photon energy. 
When $\tau_{\rm m}$ is relatively large ($\sim 0.01$), 
  a significant fraction of accreting material is lost 
  from the post-shock region at large height (see Fig.~1). 
Because we require all of the accretion energy to be liberated 
  before the matter diffuses out of the post-shock region, 
  the accretion energy is radiated at higher energies 
  and the 1--20~keV region of the resulting spectrum is enhanced 
  with respect to that of the spectrum 
  from the conventional post-shock region.  
 
For 1.0-$\Msun$ white dwarfs 
  the spectrum below $\sim 5$ keV is harder than 
  that from the conventional post-shock region, 
  while above this value it is softer. 
This effect is enhanced if significant heat flux is permitted 
  from the white dwarf (larger values of $\tau_{\rm h}$). 
For lower values of $\tau_{\rm m}$, the temperatures are already low 
  at the base of the region, 
  and the main effect is that of the increased heating from the white dwarf 
  at energies below 1~keV. 
The spectrum in this case is softer than 
  that from the conventional formulation, but only at low energies. 
When both $\tau_{\rm m}$ and $\tau_{\rm h}$ tend to zero, 
  the spectrumretains the conventional form.   
   
\subsection{Observational consequences}  

From Figure~5 it is clear that 
  it is possible to obtain either a harder or softer spectrum than 
  that from the conventional post-shock region, 
  depending on leakage parameter $\tau_{\rm m}$ 
  and heating parameter $\tau_{\rm h}$, 
  and on the energy range considered. 
When fitting to X-ray spectral data, 
  a harder model spectrum will result in a lower-mass determination 
  for the white dwarf. 
If $\tau_{\rm m}$ is significant, 
  then for typical CCD or proportional counter detectors 
  operating in the 0.2--20~keV range, 
  fitting spectra from the ``leaky'' formulation 
  will generally therefore result in lower masses for more massive white dwarfs 
  than those obtained using the Aizu (1973) or Wu et\,al.\ (1994) formulations. 
In the case of lower-mass white dwarfs the effect 
  is to mimic the change in spectral slope 
  resulting from increased cyclotron cooling from a strong magnetic field.

\begin{figure}
\vspace*{0.4cm} 
\begin{center}
\begin{tabular}{ccc}
  \epsfxsize=7cm \epsfbox{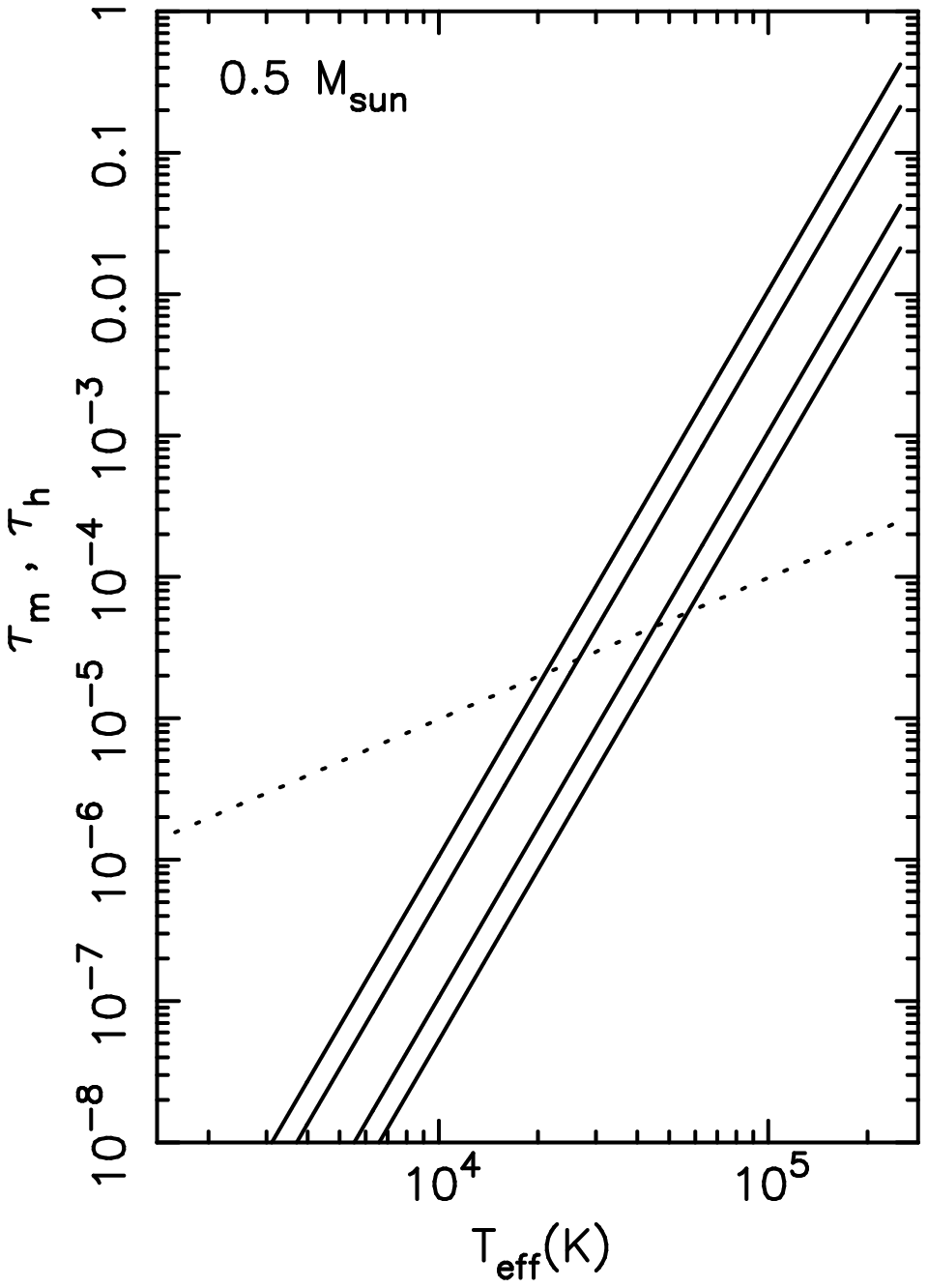} & 
  \epsfxsize=7cm \epsfbox{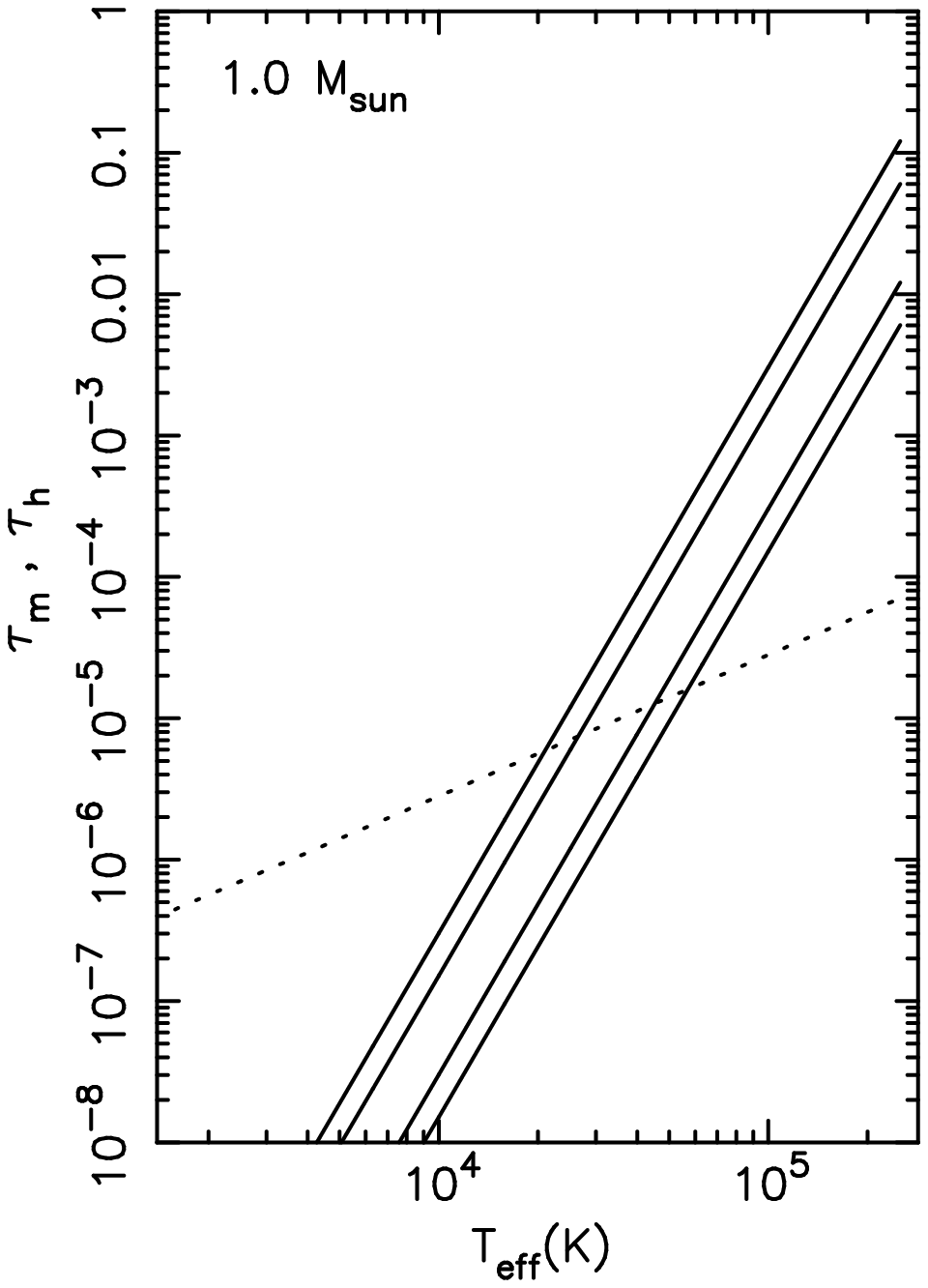}  
\end{tabular} 
\end{center}
\caption{
The values of $\tau_{\rm m}$ (dotted lines) 
  and $\tau_{\rm h}$ (solid lines) 
  as a function of the effective temperature 
  of the white-dwarf atmosphere $T_{\rm eff}$ 
  for 0.5- and 1.0-M$_\odot$ white dwarfs 
  (left and right panel respectively). 
The values of specfic accretion rates to calculate $\tau_{\rm h}$ 
  are 0.5, 1.0, 5.0 and and 50.0~g~cm$^{-2}$s$^{-1}$ 
  (solid lines from left to right). 
Here, we have considered the mass-radius relation of white dwarfs 
  given in Nauenberg (1972) 
  and assumed that the temperature at the boundary $T(0)$ 
  is the same as $T_{\rm eff}$. 
The shock temperature $T_{\rm s}$ is calculated 
  by assuming a strong adiabatic-shock condition and a cosmic abundance. }
\end{figure} 

It should be noted that because they are parameterisations, 
  the treatments in \S 2.2 and \S 2.3 do not provide a prescription 
  for definite values of $\tau_{\rm m}$ and $\tau_{\rm h}$ 
  based on atomic physics and magneto-hydrodynamics. 
However, if we simply assume certain values 
  for the effective temperature $T_{\rm eff}$ 
  of the white-dwarf atmosphere, 
  then we can obtain an estimate for $\tau_{\rm h}$ by equation (26) 
  after the specific accretion rates 
  (or the specific pre-shock mass flux) $\dot m$ 
  and the white-dwarf mass $M_{\rm w}$ are specified. 
If we further assume that the temperature $T(0)$ at the boundary 
  (i.e.\ where $\tau$ and its derivative is zero) 
  as the unperturbed effective temperature of the white-dwarf atmosphere, 
  then we may also obtain an estimate for $\tau_{\rm m}$ 
  using equation (31). 
In reality, we expect $T(0)$ to be larger because of radiative heating 
  by the X-rays from the post-shock region above and 
  $T_{\rm s}$ to be less than that determined by the free-fall velocity 
  at the shock surface (see Cropper et\,al.\ 1999). 
Therefore, we have underestimated $\tau_{\rm m}$ here, 
  and the value that we obtained can be considered 
  as a lower limit to $\tau_{\rm m}$. 
Nevertheless, under this simple approximation, 
  we have found that 
  the assumption $\tau_{\rm h} < \tau_{\rm m}$ is acceptable, 
  provided that $T_{\rm eff}$ is not very significantly larger than $10^4$~K. 
Moreover, the assumption that $\tau_{\rm m} \ll 1$ 
  is in general valid. This is shown in Figure~6. 

Which values of $\tau_{\rm m}$ can be reached is therefore unclear 
  pending further investigation, 
  but the values indicated from Figure~6 are 
  in the range $10^{-4}$ to $10^{-5}$, 
  significantly lower than those used in Figure~5. 
The difference between spectra from models 
  with these smaller values of $\tau_{\rm m}$ 
  and those from the Aizu (1973) formulation 
  is less than 1\% in the range of 0.05~eV to 100~keV, 
  so that the observational effects of the leaky base 
  will benegligible. 
Thus if we can indeed rely on small values of $\tau_{\rm m}$, 
  current cold wall models such as those mentioned in Section~1 
  can continue to be employed for X-ray spectral fitting.

The advantage of our treatment is that it avoids the infinities 
  at the base of the flow inherent in all previous formulations 
  in which the stationary-wall boundary condition base is assumed 
  (e.g.\ Aizu 1973; Chevalier \& Imamura; Wu 1994; Wu et\,al.\ 1994). 
This is a step in the direction of more realistic formulations 
  which allow the hydrodynamic accretion flow 
  to match the hydrostatic white-dwarf atmosphere. 
As noted above the predicted spectral changes are small 
  for small $\tau_{\rm m}$; 
  however a {\em practical} advantage is that 
  it eliminates the numerical errors resulting 
  from finite sampling of quantities 
  which tend to infinity. 
In the previous treatments the fineness of the sampling 
  of the base of the post-shock region 
  in the numerical integrating scheme 
  has an effect on the spectral slope: 
  finer sampling will encounter a higher value of the density 
  (which is tending to infinity) at the base, 
  so that softer spectra will be generated. 
In our formulation, as the singularity at the base is removed, 
  the effect of the sampling is insignificant 
  even when small values of $\tau_{\rm m}$ are assumed. 
This is consistent with the fact that 
  emission from the boundary layer should be finite, 
  as deduced from equation (45).

\section{ACKNOWLEDGEMENTS}

We thank Mark Wardle for comments and for carefully reading the manuscript. 
We also thank the referee for providing helpful insights for Figure~5 
  and the editor for additional comments. 
KW acknowledges the support of an ARC Australian Research Fellowship 
  and a PPARC visiting fellowship.

\appendix  

\subsection*{A.1 Convergence of the power series $\sum C_nx^n$}
  
Consider a power series  
\begin{eqnarray}  
 f(x) & = & \sum_{n=0}^{\infty} C_n x^n\  \hspace*{1cm} (x<1), \nonumber 
\end{eqnarray}  
where 
\begin{eqnarray}  
 C_n & = & (-1)^n \bigg[  {1\over {(n+1)!}} 
   - {1 \over {(n+2)!}}  \bigg]\ .  \nonumber 
\end{eqnarray}   
The series is convergent for $x <1$ if the ratio of the coefficients 
 $|C_{n+1}|/|C_{n}| \leq 1$ for sufficiently large $n$. As  
\begin{eqnarray}  
 {\lim_{n \to \infty}}~ {{|C_{n+1}|} \over {|C_n|}} 
   & = & {\lim_{n \to \infty}}~ {{(n+2)} \over {(n+1)(n+3)}} \nonumber \\ 
   & = & 0   \   ,             \nonumber 
\end{eqnarray}    
the convergent radius of $f(x)$ is infinite. Moreover, for $x>0$
\begin{eqnarray}  
 |f(x)| & \leq & \sum_{n=0}^{\infty} |C_n| x^n\ \nonumber \\ 
     & = &  \sum_{n=0}^{\infty} {1 \over {(n+2)}} 
        {x^n \over {n!}} \nonumber \\ 
     & < &  e^x \ .             \nonumber  
\end{eqnarray}
Hence, $|f(x)| < 1$ for $x \rightarrow  0^{^+}$.
 
\subsection*{Appendix A.2 Accretion luminosity}  

For the ``leaky'' accretion column with a specific mass flux 
$\sigma = \mu (1-e^{-\tau/\tau_{\rm m}})$, where 
$\mu = (1-e^{-1/4\tau_{\rm m}})^{-1}$, the accretion luminosity is 
\begin{eqnarray} 
L_{\rm acc}(\xi,1) & = & {1 \over {\gamma -1}}\int_{\tau(\xi)}^{1/4} 
    d\tau  \ 
    \biggl[\gamma - (\gamma+1)~\tau \sigma - 
    {1 \over 2} (\gamma +1)~\tau^2 
    {{d\sigma}\over{d\tau}} \biggr] \nonumber \\ 
 & = & {1 \over {\gamma -1}}\int_{\tau(\xi)}^{1/4} d\tau  \ 
    \biggl[\gamma - \mu(\gamma+1) \tau + \mu(\gamma+1) 
    \tau~e^{-\tau/\tau_{\rm m}}  
    -{1 \over 2}(\gamma +1)  \biggl({{\tau^2}
    \over{\tau_{\rm m}}}\biggr)~e^{-\tau/\tau_{\rm m}} 
    \biggr]  \nonumber \\ 
 & = &  {1 \over {\gamma -1}} \big[ I_1(\xi,1) - 
    I_2(\xi,1) + I_3(\xi,1) -I_4(\xi,1) \big] \nonumber \ , 
\end{eqnarray}    
where 
\begin{eqnarray}  
 I_1(\xi,1) & = & \gamma\ \int_{\tau(\xi)}^{1/4} d\tau  \nonumber  \\ 
     & = & \gamma~ \biggl[ {1 \over 4} -\tau \biggr] \ ,  \nonumber \\ 
 I_2(\xi,1) & = & \mu(\gamma+1) \int_{\tau(\xi)}^{1/4} 
    d\tau \ \tau \nonumber \\ 
     & = & {\mu \over 2} (\gamma+1)~  
    \biggl[ {1 \over 16} -\tau^2 \biggr]
     \ , \nonumber  \\ 
 I_3(\xi,1) & = & \mu (\gamma+1) \int_{\tau(\xi)}^{1/4} 
     d\tau \ \tau~e^{-\tau/\tau_{\rm m}} \nonumber \\  
      & = & \mu (\gamma+1) \tau_{\rm m}^2~ \biggl[
     \biggl( 1 + {\tau \over {\tau_{\rm m}}} \biggr)  
    e^{-\tau/\tau_{\rm m}} - 
    \biggl( 1 + {1\over {4\tau_{\rm m}}} \biggr) e^{-1/4\tau_{\rm m}} \biggr] 
   \ , \nonumber \\ 
 I_4(\xi,1) & = & {\mu \over 2} (\gamma+1) \int_{\tau(\xi)}^{1/4} 
   d\tau \ 
        {{\tau^2}\over {\tau_{\rm m}}}~e^{-\tau/\tau_{\rm m}}  \nonumber \\   
    & = & \mu (\gamma+1) \tau_{\rm m}^2~ \biggl[ 
       \biggl( 1 + {\tau \over {\tau_{\rm m}}} + 
    {\tau^2 \over {2 \tau_{\rm m}^2}}  \biggr) e^{-\tau/\tau_{\rm m}}
          \biggl( 1 + {1\over {4\tau_{\rm m}}} + 
     {1\over {32\tau_{\rm m}^2}}\biggr) e^{-1/4\tau_{\rm m}} \biggr] \ .  
    \nonumber    
\end{eqnarray}  
Hence, we have  
\begin{eqnarray} 
L_{\rm acc}(\xi,1) & = &   {1 \over {\gamma -1}}  
     \biggl\{ {\gamma \over 4}  - {\mu \over 32} (\gamma+1)  -\gamma\tau 
    +  {\mu \over 2} (\gamma+1) \tau^2 
     + \mu (\gamma+1) \tau_{\rm m}^2 \biggl[  
     \biggl( 1 + {\tau\over {\tau_{\rm m}}} \biggr)
        -  \biggl(1 + {\tau\over {\tau_{\rm m}}}+ 
    {\tau^2 \over {2\tau_{\rm m}^2}} \biggr) 
      \biggr] e^{-\tau/\tau_{\rm m}}   \nonumber \\ 
     &  & \ \ \  
   - \mu (\gamma+1) \tau_{\rm m}^2 \biggl[ 
   \biggl( 1 + {1\over {4\tau_{\rm m}}} \biggr) 
    - \biggl( 1 + {1\over {4\tau_{\rm m}}}  + 
      {1\over {32\tau_{\rm m}^2}}\biggr) \biggr] e^{-1/4\tau_{\rm m}}    
      \biggr\}    \nonumber \\  
     & = & {1 \over {\gamma -1}} \bigg\{ {\gamma \over 4} 
        - {\mu \over 32} (\gamma+1)~\big[1- e^{-1/4\tau_{\rm m}} \big] 
          -\gamma \tau \biggl[ 1 - {{\mu (\gamma+1)} \over {2\gamma}}  
        \big(1 - e^{-\tau/\tau_{\rm m}} \big)\tau  \biggr] \bigg\} \nonumber \\ 
     & = & {1 \over {\gamma -1}} \bigg\{ {\gamma \over 4} 
        - {1 \over 32} (\gamma+1)          
       -\gamma \tau \biggl[ 1 - {{\gamma+1} \over {2\gamma}}  
        \biggl({{1 - e^{-\tau/\tau_{\rm m}}} \over {1 - e^{-1/4\tau_{\rm m}}}} 
      \biggr)\tau  \biggr] \bigg\} \ . \nonumber 
\end{eqnarray}    

\end{document}